%% file: main.tex
\documentclass[twocolumn,prd, aps, notitlepage, nofootinbib, showpacs, superscriptaddress,longbibliography]{revtex4-1}
\usepackage{amssymb,amsmath,amsfonts,graphicx,epsf,soul}
\usepackage{aas_macros}
\usepackage{lipsum}
\usepackage[dvipsnames]{xcolor}
\usepackage{bm}
\usepackage{listings}
\definecolor{codegreen}{rgb}{0,0.6,0}
\definecolor{codegray}{rgb}{0.5,0.5,0.5}
\definecolor{codepurple}{rgb}{0.58,0,0.82}
\definecolor{backcolour}{rgb}{0.95,0.95,0.92}

\usepackage{mathrsfs}
\usepackage[scr=boondox,  
            cal=esstix]   
           {mathalpha}
           
\newcommand{\aEMRI}{\ensuremath{a_{\bullet}}}
\newcommand{\mninefive}{\ensuremath{m_{95}}}
\newcommand{\mEMRI}{\ensuremath{m_{\bullet}}}
\newcommand{\pdisk}{\ensuremath{\mathfrak{p}}}
\newcommand{\qdisk}{\ensuremath{\mathfrak{q}}}
\newcommand{\OmegaEMRI}{\ensuremath{\Omega_{\bullet}}}
\newcommand{\epssoft}{\ensuremath{\epsilon_{\mathrm{soft}}}}
\newcommand{\bsoft}{\ensuremath{b}}

\makeatletter
\newcommand{\oset}[3][0ex]{%
  \mathrel{\mathop{#3}\limits^{
    \vbox to#1{\kern-2\ex@
    \hbox{$\scriptstyle#2$}\vss}}}}
\makeatother

\graphicspath{{./}{figures/}}

\usepackage{hyperref}%
\hypersetup{
  colorlinks=true,       %
  linkcolor=MidnightBlue,%
  citecolor=MidnightBlue,%
  urlcolor=MidnightBlue  %
}

\allowdisplaybreaks[4]
\begin{document}

\title{Extreme-Mass-Ratio Inspirals in Gaseous Disks}

\author{Alexander J. Dittmann}
\altaffiliation{NASA Hubble Fellowship Program, Einstein Fellow}
\affiliation{Institute for Advanced Study, 1 Einstein Drive, Princeton, NJ 08540, USA}
\thanks{The first and second authors contributed equally.}

\author{Abhishek Hegade K. R.}
\affiliation{Princeton Gravity Initiative, Princeton University, Princeton, NJ 08544, USA}

\author{Callum W. Fairbairn}
\affiliation{Institute for Advanced Study, 1 Einstein Drive, Princeton, NJ 08540, USA}

\begin{abstract} 
Gravitational waves from extreme mass ratio inspirals (EMRIs) are precise probes of the environment of the supermassive black holes (SMBHs) they orbit.
If an SMBH is actively accreting, the surrounding gaseous disk can impart hydrodynamic torques on and assist the formation of EMRIs within it.
Such disk-EMRI interactions could leave measurable imprints on future observations by the Laser Interferometer Space Antenna (LISA), and potentially provide a route to constrain disk properties using gravitational wave observations.
We present herein a detailed relativistic analysis of these hydrodynamic interactions using linear theory. We first derive a Lagrangian governing the evolution of spiral density waves in the disk and use it to formulate a balance law for the transfer of angular momentum between the EMRI and disk.
We then develop a stable numerical scheme which can be used to treat corotation resonances and find modal solutions in thin disks up to very large azimuthal numbers.
Using this numerical scheme, we explore how SMBH spins, EMRI semi-major axes, disk scale heights, sound speed gradients, and surface density gradients affect the interaction between accretion disks and circular EMRIs.
Our results show that relativistic effects substantially alter disk-EMRI interactions once the secondary orbit is within $\mathcal{O}(25)$ Schwarzschild radii from the SMBH. 
Comparing our numerical results with recent analytical models suggests that the impact of pressure gradients and softening of the gravitational potential is important for disks with finite thickness and cannot be captured by tuning the torque cutoff parameters in the analytical models.
The framework provided here will help analyze the formation scenarios of EMRIs and build relativistically accurate waveform models of disk-EMRI interactions.
\end{abstract}
\maketitle
\section{Introduction}
Black hole binaries with extreme mass ratios, comprised of a supermassive black hole (SMBH) and a stellar-mass black hole, are thought to occur in the centers of virtually all large galaxies \citep[e.g.][]{2013ARA&A..51..511K,2017ARA&A..55...17A,2020A&ARv..28....4N}. 
The gravitational waves emitted by these extreme mass ratio inspirals (EMRIs) hold great potential to probe gravitational physics in the strong-field regime, and are one of the main targets for the Laser Interferometer Space Antenna (LISA) and other space-based gravitational wave (GW) detectors~\cite{LISA:2022yao,Luo_2016}. Because these sources can undergo many thousands of cycles within a few-year observational window, their gravitational waveforms may accumulate measurable imprints of deviations from vacuum general relativity, be those from astrophysical environments~\cite{Kocsis_2011,Yunes:2011ws,Silva:2022blb,Yin_2025,Silva:2025lkl,Lui:2026uai}, exotic sources such as dark matter spikes~\cite{Speeney:2022ryg,Vicente_2022,Mitra:2025tag,Vicente:2025gsg} and scalar clouds~\cite{Barsanti:2022ana,Speri:2024qak,dyson2025environmentaleffectsextrememass,Li:2025ffh,Xu:2026cky,Xu:2026aic}, or potential corrections to general relativity in the strong-field regime~\cite{Yang:2026erj,Zi:2026zpw,Gliorio:2026yvh}.

Approximately ten percent of SMBHs actively accrete from gaseous disks at a given time \citep{2017A&A...601A..63W,2024MNRAS.527.3006G},
naively suggesting that a similar fraction of EMRIs will take place in gaseous environments. Moreover, such gaseous environments could enhance EMRI rates~\cite{Pan_2021,Zeng:2026ydj}, both by capturing stellar-mass black holes and providing a sink for their orbital energy, and potentially by forming stars and black holes in situ via gravitational instability. 
In one view, not accounting for the presence of these gaseous disks might bias tests of general relativity~\cite{Barausse:2014tra,Copparoni:2025vty,Kejriwal:2025jao} and downstream EMRI physics; more positively, EMRIs will provide us with a window into the inner regions of these accretion flows~\cite{Speri_2023,duque2025constrainingaccretionphysicsgravitational,Fantoccoli:2026idl}, the physical conditions of which are still highly uncertain~\cite{2014ARA&A..52..529Y,Blaes_2013,Abramowicz_2013}.

Early studies of disk-EMRI interactions drew upon a rich history of work on planet-disk interactions in the Newtonian regime thanks to the similarity in mass ratios between the systems~\cite{Yunes:2011ws,Kocsis_2011}. Relevant works in the Newtonian regime include analytical estimates of the torques on disk-embedded objects in the WKB limit~\cite{1979ApJ...233..857G,GT-disc-satellite-interaction,Ward-1986,1993ApJ...419..155A,Ward-1997,2002ApJ...565.1257T}, numerical solutions of the linearized fluid equations~\cite{K&P_1993,Miranda_2019,Miranda_2020}, and nonlinear hydrodynamical simulations~\citep{Paardekooper:2023,Derdzinski:2025cql}. 
While these calculations provide valuable insights into the nature of the disk-EMRI interaction, the applicability of these calculations to the inner regions of SMBH accretion disks is limited by a number of factors, such as their assumption of a disk dominated by gas pressure rather than radiation pressure, their assumption of a laminar rather than turbulent environment, and neglect of relativistic effects irrelevant in star-planet binaries but of crucial importance to EMRIs in the LISA band.

Recently, several studies have focused on extending the theory of disk-EMRI interactions into the relativistic regime for EMRIs on circular orbits. Techniques such as Hamiltonian perturbation theory~\cite{HegadeKR:2025rpr,HegadeKR:2025dur}, black hole perturbation theory~\cite{Hirata_2011,Hirata_2011_II,Duque:2025yfm} and numerical integrations of linearized fluid equations about Kerr spacetime~\cite{Dyson:2026ddd} have been used to assess the importance of the strong-field gravity of the SMBH on the disk-EMRI interactions. 
Analytical~\cite{HegadeKR:2025rpr,HegadeKR:2025dur} and numerical results~\cite{Duque:2025yfm,Dyson:2026ddd} have suggested that relativistic effects enhance the magnitude of the torque close to the innermost stable orbit (ISCO) where EMRIs in the LISA band are expected to be detected.
In the case of thin-pressureless disks, it has also been suggested that the direction of the gas-induced torque on the EMRI might reverse as it approaches the inner edge of the accretion disk~\cite{HegadeKR:2025rpr,HegadeKR:2025dur,Duque:2025yfm}. 

To assess these predictions and consistently incorporate finite-pressure fluid dynamics, one requires a formalism capable of determining the full linear response of a relativistic disk. In this paper, we develop such a framework by constructing a linear perturbation theory for stationary and axisymmetric relativistic accretion disks. Adapting techniques from the theory of rotating relativistic  stars~\cite{FN-Book}, we derive a Lagrangian formulation of the perturbation equations together with energy- and angular-momentum-balance laws governing the interaction between the disk and the EMRI. We use this framework to calculate the linear disk-induced torque across a broad range of disk profiles, SMBH spins, orbital radii, and disk scale heights, and to determine the morphology of the associated spiral-density response. In particular, we show how finite pressure sets the high-$m$ cutoff of the torque spectrum, thereby determining the number of azimuthal modes required for accurate numerical calculations and providing guidance for setting the torque cutoff parameter in analytical models~\cite{HegadeKR:2025dur,HegadeKR:2025rpr,Duque:2025yfm}.

Carrying out these calculations requires a stable numerical method capable of resolving thin, low-pressure disks, treating the singular response near corotation, and efficiently summing over a large number of azimuthal modes. A recent calculation demonstrated the feasibility of numerically determining the relativistic linear response~\cite{Dyson:2026ddd}; 
that study included the full metric perturbation of the EMRI, but was limited comparatively thick disks and approximately $15$ azimuthal modes,\footnote{As we demonstrate in Sec.~\ref{sec:results}, it is necessary to calculate the response of the disk up to $\sim 3/h$ azimuthal modes to achieve a $\sim5\%$ accurate calculation of the total torque, where $h$ is the aspect ratio of the disk.} and was limited to disks where the corotation resonance was analytically integrable. 
We address these difficulties by developing an efficient implicit numerical scheme that integrates two first-order equations for the enthalpy and the radial velocity perturbation instead of the conventional, single second-order enthalpy equation.
We implement a Landau-type regularization of the corotation singularity and construct an adaptive radial grid that resolves both the corotation layer and the highly oscillatory WKB-like spiral-density waves away from the Lindblad resonances. This allows us to compute the large number of azimuthal modes required for converged torque estimates in thin disks.

The paper is organized as follows. In Sec.~\ref{sec:linear-theory}, we develop the relativistic perturbation theory, derive the energy- and angular-momentum-balance laws, and describe our numerical implementation and treatment of corotation. In Sec.~\ref{sec:results}, we present the spiral-density response and disk-induced torques, examine their dependence on relativistic effects and disk properties, and compare our results with pressureless disk models. We conclude in Sec.~\ref{sec:conclusions}. Throughout the paper, we use geometric units with $G=c=1$ and the metric signature $(-,+,+,+)$.
\section{Perturbation theory for spiral density waves}\label{sec:linear-theory}
In this section, we develop the relativistic linear theory of spiral density waves raised by the motion of EMRIs in a thin, vertically-integrated, two-dimensional, steady, and axisymmetric accretion disk. 
The general perturbation equations --- valid in full three-dimensional self-gravitating disks --- have been developed long ago in~\cite{FS_stability_rel,FS-Non-Relativitistic,Ipser-Lindblom-1992,FN-Book}. Our formalism relies heavily on these references, and a detailed introduction to the tools used in these papers can be found in Chapter 7 of~\cite{FN-Book}.

The rest of this section is organized as follows.
In Sec.~\ref{sec:eom-and-bkg}, we describe the basic equations and the background axisymmetric disk structure.
Section~\ref{sec:lagrangian-perturbation-theory} presents the details of Lagrangian perturbation theory and provides the definition of the conserved energy and angular momentum of spiral density waves.
We then describe our master equations in Sec.~\ref{sec:master-eqs} and present the simplified gravitational potential of the secondary compact object used in the paper in Sec.~\ref{sec:EMRI-potential}.
Section~\ref{sec:numerics} contains the details of our numerical implementation and the diagnostics used to obtain the torque on the disk.
\subsection{Equations of motion and background structure}\label{sec:eom-and-bkg}
Consider a razor thin accretion disk around a background SMBH spacetime $g_{\mu\nu}$ modeled as an inviscid fluid of mass density $\rho$ and stress-energy tensor
\begin{align}
    T_{\mu\nu} &= \varepsilon u_{\mu} u_{\nu} + p q_{\mu \nu}
\end{align}
where $u^{\mu}$ is the fluid four-velocity, $p$ is the height-integrated pressure, $\varepsilon$ is the height integrated energy density and 
\begin{align}
    q_{\mu\nu} \equiv u_{\mu} u_{\nu} + g_{\mu\nu} \,,
\end{align}
is the projection-tensor.
The equations of motion of the disk are the mass conservation equation, the energy conservation equation and the Euler equation
\begin{subequations}\label{eq:full-eqn-set-source-zero}
\begin{align}
\label{eq:continuity-eqn}
    &u^{\mu}\nabla_{\mu} \rho + \rho \nabla_{\alpha} u^{\alpha}= 0 \,,\\
\label{eq:energy-eqn}
    &u^{\mu}\nabla_{\mu} \varepsilon  + (\varepsilon + p) \nabla_{\alpha} u^{\alpha} = 0\,,\\
\label{eq:euler-eqn}
    &\left(\varepsilon + p \right) u^{\nu}\nabla_{\nu} u^{\mu}  
    + 
    q^{\mu \alpha} \nabla_{\alpha}p = 0 \,.
\end{align}
\end{subequations}

We assume that the background spacetime of the SMBH is stationary, axisymmetric, asymptotically flat and reflection symmetric around the equatorial plane.
With these assumptions, one can decompose the metric close to the equatorial plane in a coordinate system $(t,r,z,\phi)$ such that~\cite{Page-Thorne-1974}
\begin{align}\label{eq:qI-line-element}
    ds^2 &= - e^{2\nu} dt^2 + e^{2 \psi} \left(d \phi - \tilde{\omega} dt \right)^2 + e^{2 \mu} d r^2 + dz^2\,,
\end{align}
where the metric functions $\nu,\psi,\tilde {\omega}$ and $\mu$ only depend on $r$. The explicit form of the metric functions for a Kerr spacetime are provided in Appendix~\ref{appendix:Kerr-In-Plane-Coeffs}.

The fluid equilibrium state is assumed to be stationary, axisymmetric and purely rotational with a 4-velocity given by
\begin{align}\label{eq:4-velocity-decomposition}
    u^{\alpha} &= u^t \left( t^{\alpha} + \Omega \phi^{\alpha}\right)
\end{align}
where $t^{\alpha}$ and $\phi^{\alpha}$ are the stationary and axisymmetric Killing vectors in the spacetime and $\Omega$ is the rotational velocity of the fluid.
With these assumptions, one can show that
\begin{align}
    \nabla_{\alpha} u^{\alpha} = 0\,. 
\end{align}
Therefore, Eqs.~\eqref{eq:continuity-eqn} and \eqref{eq:energy-eqn} are readily satisfied.
The Euler equation [Eq.~\eqref{eq:euler-eqn}] can be simplified as
\begin{align}
    a_{\alpha} \equiv u^{\mu} \nabla_{\mu} u_{\alpha}
    =
    u^t u_{\phi} \nabla_{\alpha} \Omega -\nabla_{\alpha} \log u^t 
    = - \frac{\nabla_{\alpha} p}{(\varepsilon + p)}
    \,.
\end{align}
To simplify the above equation further, we decompose the 4-velocity as
\begin{align}
    &u^t = \frac{e^{-\nu}}{\sqrt{1-v^2}}\,, &u^{\phi}& = \Omega u^t \,,\\
    &u_{t} =-\frac{e^{\nu}}{\sqrt{1-v^2}} \left(1 + e^{\psi - \nu} \tilde{\omega} v \right) \,, &u_{\phi}& = \frac{e^{\psi} v}{\sqrt{1-v^2}} \,,
\end{align}
where
\begin{align}\label{eq:v-def}
    v \equiv (\Omega - \tilde{\omega}) e^{\psi - \nu} \,.
\end{align}
Using this decomposition we can simplify the Euler equation to~\cite{FN-Book}
\begin{align}\label{eq:dpdr-equilibrium}
    \frac{ p'}{\varepsilon + p}
    &=
    -\frac{1}{1-v^2} \left(\nu' - v^2 \psi' + e^{\psi - \nu} v \tilde{\omega}' \right)
    \,,
\end{align}
where $'$ denotes differentiation with respect to $r$.
Given $g_{\mu\nu}$, $\varepsilon$, and $p$ one can obtain $v(r)$ and $\Omega(r)$ using Eq.~\eqref{eq:dpdr-equilibrium} and generate rotational velocity of equilibrium configuration from Eq.~\eqref{eq:v-def}.

In this paper, we use simple power laws to represent the equation of state of the background fluid 
\begin{subequations}
\begin{align}
    &\varepsilon = \varepsilon_0 \left( \frac{r}{\aEMRI} \right)^{-\pdisk} \,,\\
    &p = c_s^2 \varepsilon \,,\\
    &c_s = h_0\, \aEMRI\, \OmegaEMRI(\aEMRI)\, \left( \frac{r}{\aEMRI} \right)^{-\qdisk/2}
    \,,
\end{align}
\end{subequations}
where $\varepsilon_0$ is a fiducial energy density, $h_0$ is a fiducial scale height, $(\pdisk,\qdisk)$ are free constants, $c_s$ is the sound speed, $\aEMRI$ is the semi-major axis of the EMRI and $\OmegaEMRI(\aEMRI)$ is the orbital frequency of the EMRI. For a Kerr SMBH of mass $M$ and dimensionless spin $\chi$
\begin{align}\label{eq:OmegaEMRIKerr}
    \OmegaEMRI(\aEMRI) = \frac{1}{M \left[(\aEMRI/M)^{3/2} + \chi \right]}
    \,.
\end{align}

\subsection{Lagrangian perturbation theory and conserved quantities}\label{sec:lagrangian-perturbation-theory}
We now perturb the equations of motion of the fluid [  Eq.~\eqref{eq:full-eqn-set-source-zero}] to account for the forcing produced by the EMRI. Let $\delta$ denote an Eulerian perturbation and 
\begin{align}
h_{\mu\nu} \equiv \delta g_{\mu\nu}
\end{align}
denote the linear metric perturbation sourced by the EMRI; see Sec.~\ref{sec:EMRI-potential} below for exact expressions. Linearizing the mass conservation, energy conservation, and Euler equations leads to
\begin{subequations}\label{eq:perturbed-system}
\begin{align}
&\delta \left[
u^{\mu}\nabla_{\mu} \rho
+
\rho \nabla_{\alpha} u^{\alpha}
\right] = 0,
\label{eq:perturbed-system-mass}
\\
&\delta\left[u^{\mu}\nabla_{\mu} \varepsilon  + (\varepsilon + p) \nabla_{\alpha} u^{\alpha}\right] = 0,
\label{eq:perturbed-system-energy}
\\
&\delta\left[
\left(\varepsilon+p\right)a_{\alpha}
+
q_{\alpha}{}^{\lambda}\nabla_{\lambda}p
\right]=0,
\label{eq:perturbed-system-euler}
\end{align}
\end{subequations}

To simplify the perturbation equations, it is useful to introduce a Lagrangian displacement vector $\xi^{\alpha}$. The Eulerian velocity perturbation is related to $\xi^\alpha$ by~\citep{FN-Book}
\begin{align}
\delta u^{\alpha}
=
q^{\alpha}{}_{\beta}\mathcal{L}_{u}\xi^{\beta}
+
\frac{1}{2}
u^{\alpha}u^{\beta}u^{\gamma}h_{\beta\gamma},
\label{eq:delta-u-displacement}
\end{align}
where $\mathcal{L}$ denotes the Lie derivative. The corresponding Lagrangian perturbation $\Delta$ is
\begin{align}
\Delta \equiv \delta+\mathcal{L}_{\xi}\,.
\end{align}
In particular,
\begin{align}
\Delta g_{\alpha\beta}
=
h_{\alpha\beta}
+
2\nabla_{(\alpha}\xi_{\beta)}\,.
\label{eq:Delta-g-def}
\end{align}
Using Lagrangian perturbations, the perturbed mass conservation equation and energy conservation can be solved in terms of the Lagrangian perturbation of the metric
\begin{align}
\frac{\Delta\rho}{\rho}
=
\frac{\Delta\varepsilon}{\varepsilon+p}
=
-\frac{1}{2}
q^{\alpha\beta}\Delta g_{\alpha\beta}\,.
\label{eq:Delta-rho-epsilon}
\end{align}
To solve for the pressure perturbations, we assume that the pressure perturbations are adiabatic with adiabatic index $\Gamma_1$, so that
\begin{align}
\frac{\Delta p}{p}
=
\Gamma_1\frac{\Delta\rho}{\rho}
=
\Gamma_1\frac{\Delta\varepsilon}{\varepsilon+p}\,.
\label{eq:Delta-p-adiabatic}
\end{align}
Equivalently,
\begin{align}
\Delta p
=
-\frac{1}{2}
\Gamma_1 p
\,
q^{\alpha\beta}\Delta g_{\alpha\beta}\,.
\label{eq:Delta-p-explicit}
\end{align}
In this paper, we pick a simple parameterization for the adiabatic index
\begin{align}
    \Gamma_1 = \frac{\varepsilon + p}{p} c_s^2
    \,.
\end{align}
This parameterization affects the numerical results presented in Sec.~\ref{sec:results} but it does not impact any of the symbolic derivations presented here.

We now focus on the Euler equation in order to understand how the EMRI exchanges energy and angular momentum with the disk. Define the unperturbed Euler equation schematically as
\begin{align}
\mathcal{E}_{\alpha}
\equiv
\left(\varepsilon+p\right)a_{\alpha}
+
q_{\alpha}{}^{\lambda}\nabla_{\lambda}p\,.
\label{eq:euler-operator-background}
\end{align}
The background satisfies
\begin{align}
\mathcal{E}_{\alpha}=0\,.
\end{align}
By eliminating the mass, energy and pressure perturbations using Eqs.~\eqref{eq:Delta-rho-epsilon} and \eqref{eq:Delta-p-explicit}, we can express the linearized Euler equation entirely in terms of the Lagrangian metric perturbation $\Delta g_{\alpha\beta}$ [see, Eq.~(7.16) of~\cite{FN-Book}]:
\begin{align}
&\delta \mathcal{E}_{\alpha}
=
\left(\varepsilon + p\right) \mathcal{L}_{u} \left(q_{\alpha}{}^{\beta} u^{\gamma} \Delta g_{\beta \gamma} \right) 
\nonumber\\ 
&- 
\frac{1}{2} \left(\varepsilon + p \right)q_{\alpha}{}^{\beta} \nabla_{\beta}\left(u^{\gamma} u^{\delta} \Delta g_{\gamma \delta} \right) 
\nonumber\\ 
&+ \frac{1}{2} \left(1 + \frac{\Gamma_1 p}{\varepsilon + p}\right) q^{\gamma \delta} \Delta g_{\gamma \delta} \nabla_{\alpha} p \nonumber\\ 
&- \frac{1}{2} q_{\alpha}{}^{\beta} \nabla_{\beta}\left(\Gamma_1 p q^{\gamma \delta} \Delta g_{\gamma \delta}\right) \,.
\label{eq:linearized-euler-Delta-g}
\end{align}
Let us denote this equation schematically as
\begin{align}
\delta\mathcal{E}_{\alpha}[\xi,h]=0\,.
\end{align}
Since $\Delta g_{\alpha\beta}$ [Eq.~\eqref{eq:Delta-g-def}] is linear in both $\xi^\alpha$ and $h_{\alpha\beta}$, we can split the equation into a homogeneous fluid-displacement part and an externally forced part:
\begin{align}
\delta\mathcal{E}_{\alpha}[\xi,h]
=
\delta\mathcal{E}_{\alpha}[\xi,0]
+
\delta\mathcal{E}_{\alpha}[0,h]\,.
\end{align}
We define
\begin{align}
E_{\alpha}[\xi]
\equiv
\delta\mathcal{E}_{\alpha}[\xi,0],
\qquad
F_{\alpha}[h]
\equiv
-
\delta\mathcal{E}_{\alpha}[0,h]\,.
\label{eq:E-and-F-def}
\end{align}
The perturbed Euler equation then takes the forced form
\begin{align}
E_{\alpha}[\xi]
=
F_{\alpha}[h]\,.
\label{eq:perturbation-eq-forced}
\end{align}
Here $E_{\alpha}$ is the fluid evolution operator obtained by setting the external metric perturbation to zero, $h_{\alpha\beta}=0$, while $F_{\alpha}$ is the force density induced by the EMRI metric perturbation, obtained by setting the fluid displacement to zero, $\xi^{\alpha}=0$.
This form makes explicit that the EMRI metric perturbation acts as an external source for the disk displacement.
The explicit expression for the force density is provided in Eq.~\eqref{eq:EMRI-Force-Analytical} below.

Using the self-adjoint structure of the relativistic fluid perturbation equations discussed in Chapter~7 of~\cite{FN-Book}, the forced equation [Eq.~\eqref{eq:perturbation-eq-forced}] can be obtained from an effective Lagrangian\footnote{Note that~\cite{FN-Book} uses Lagrangian pseudodensities. } for the displacement,
\begin{align}
\mathscr{L}_{\rm eff}[\xi;h]
=
\mathscr{L}_{\rm W}[\xi]
+
\mathscr{L}_{\rm int}[\xi,h]\,,
\label{eq:effective-forced-lagrangian}
\end{align}
where $\mathscr{L}_{\rm W}$ is the Lagrangian density governing the evolution of the free spiral density waves, $\mathscr{L}_{\rm int}$ describes the interaction between the EMRI perturbation $h_{\alpha\beta}$ and the spiral density waves. Explicitly,
\begin{subequations}
\begin{align}
&\mathscr{L}_{\rm W}
=
\frac{1}{2}
U^{\alpha\beta\gamma\delta}
\nabla_{\alpha}\xi_{\beta}
\nabla_{\gamma}\xi_{\delta}
-
\frac{1}{2}
T^{\alpha\beta}
R_{\alpha\gamma\beta\delta}
\xi^{\gamma}\xi^{\delta}\,,\\
\label{eq:cowling-quadratic-lagrangian}
&\mathscr{L}_{\rm int}
=
V^{\alpha \beta \gamma \delta}
\left(h_{\alpha \beta} \nabla_{\gamma} \xi_{\delta}
\right)
-
\frac{1}{2} 
h_{\alpha \beta} \xi^{\gamma} \nabla_{\gamma} T^{\alpha \beta}
\,,
\end{align}
\end{subequations}
where $R_{\alpha\gamma\beta\delta}$ is the background Riemann curvature, and 
\begin{align}
    &U^{\alpha \beta \gamma \delta} \equiv (\varepsilon + p) u^{\alpha} u^{\gamma} q^{\beta \delta} + p (g^{\alpha \beta} g^{\gamma \delta} - g^{\alpha \delta} g^{\beta \gamma}) \nonumber\\
    &
    - \Gamma_1 p q^{\alpha \beta} q^{\gamma \delta}
    \,,\\
    &2 V^{\alpha \beta \gamma \delta} \equiv
    \left(\varepsilon + p \right)
    \left(
    u^{\alpha} u^{\gamma} q^{\beta \delta}
    +
    u^{\beta} u^{\gamma} q^{\alpha \delta}
    - 
    u^{\alpha} u^{\beta} q^{\gamma \delta}
    \right)
    \nonumber\\
    &-
    \Gamma_1 p q^{\alpha \beta} q^{\gamma \delta}
    \,.
\end{align}
The Lagrangian formulation, Eq.~\eqref{eq:effective-forced-lagrangian},
also implies that the force due to the EMRI can be written as
\begin{align}\label{eq:EMRI-Force-Analytical}
    F^{\beta}
    =
    -
    \left[
    \nabla_{\alpha}
    \left(
        V^{\gamma \delta \alpha \lambda}
        h_{\gamma \delta}
    \right)
    +
    \frac{1}{2}
    h_{\gamma \delta}
    \nabla^{\lambda}T^{\gamma \delta}
    \right] .
\end{align}

The existence of the Lagrangian can be used to define the canonical energy $j_E^{\alpha}$ and angular momentum $j^{\alpha}_{J}$ currents of the spiral density waves
\begin{subequations}
\begin{align}
    j_{E}^{\alpha}
    &\equiv
    \Pi^{\alpha\beta}
    \mathcal{L}_{t}\xi_{\beta}
    -
    t^{\alpha}\mathscr{L}_{\rm eff},
    \\
    j_{J}^{\alpha}
    &\equiv
    -
    \Pi^{\alpha\beta}
    \mathcal{L}_{\phi}\xi_{\beta}
    +
    \phi^{\alpha}\mathscr{L}_{\rm eff},
\end{align}
\end{subequations}
where
\begin{align}
    \Pi^{\alpha\beta}
    \equiv
    \frac{\partial \mathscr{L}_{\rm eff}}
    {\partial(\nabla_{\alpha}\xi_{\beta})}
    =
    U^{\alpha\beta\gamma\delta}
    \nabla_{\gamma}\xi_{\delta}
    +
    V^{\gamma\delta\alpha\beta}
    h_{\gamma\delta}.
\end{align}
Using the forced equation of motion,
Eq.~\eqref{eq:perturbation-eq-forced}, we obtain the balance laws for the evolution of energy and angular momentum of the disk
\begin{subequations}
\begin{align}
    \nabla_{\alpha}j_{E}^{\alpha}
    &=
    \mathscr{P},
    \\
    \nabla_{\alpha}j_{J}^{\alpha}
    &=
    \mathscr{T}.
    \label{eq:angular-momentum-current}
\end{align}
\end{subequations}
The power density $\mathscr{P}$ and torque density $\mathscr{T}$ delivered to the spiral density waves
are obtained from the explicit dependence of the effective Lagrangian on the
external EMRI metric perturbation. Holding the fluid displacement fixed, we
have
\begin{subequations}
\begin{align}
    \mathscr{P}
    &=
    -
    \left.
    \mathcal{L}_{t}\mathscr{L}_{\rm eff}
    \right|_{\xi}
    \nonumber\\
    &=
    -
    V^{\alpha\beta\gamma\delta}
    \mathcal{L}_{t}h_{\alpha\beta}
    \nabla_{\gamma}\xi_{\delta}
    +
    \frac{1}{2}
    \mathcal{L}_{t}h_{\alpha\beta}
    \xi^{\gamma}\nabla_{\gamma}T^{\alpha\beta}
    \label{eq:power-1}
    ,
    \\
    \mathscr{T}
    &=
    \left.
    \mathcal{L}_{\phi}\mathscr{L}_{\rm eff}
    \right|_{\xi}
    \nonumber\\
    &=
    V^{\alpha\beta\gamma\delta}
    \mathcal{L}_{\phi}h_{\alpha\beta}
    \nabla_{\gamma}\xi_{\delta}
    -
    \frac{1}{2}
    \mathcal{L}_{\phi}h_{\alpha\beta}
    \xi^{\gamma}\nabla_{\gamma}T^{\alpha\beta}
    \label{eq:torque-1}
    \,.
\end{align}
\end{subequations}
We can simplify this equation further by using the following identity
\begin{align}
    \label{eq:identity-for-sim}
    \frac{1}{2}
    \delta_{\xi}T^{\alpha\beta}
    k_{\alpha\beta}
    =
    V^{\alpha\beta\gamma\delta}
    k_{\alpha\beta}
    \nabla_{\gamma}\xi_{\delta}
    -
    \frac{1}{2}
    k_{\alpha\beta}
    \xi^{\gamma}\nabla_{\gamma}T^{\alpha\beta},
\end{align}
where $k_{\alpha \beta}$ is any symmetric tensor and $\delta_{\xi}$ denote the Eulerian change in a quantity with $h_{\mu\nu} = 0$.
In practice, this can be computed as
\begin{align}
    \delta_{\xi} T^{\alpha \beta} = \delta T^{\alpha \beta} - \delta_{\xi=0} T^{\alpha \beta}\,.
\end{align}
The first term in the above expression can be computed from numerical solutions to the master equations and the second term is completely determined by the external potential
\begin{align}
    \delta_{\xi=0} T^{\alpha \beta} = \left[W^{\alpha \beta \gamma \delta} - \frac{1}{2} T^{\alpha \beta} g^{\gamma \delta} \right] h_{\gamma \delta}
    \,,
\end{align}
where
\begin{align}
    &W^{\alpha \beta \gamma \delta}
    \equiv
    \frac{1}{2} \left(\varepsilon + p \right) u^{\alpha}u^{\beta}u^{\gamma}u^{\delta}
    \nonumber\\
    &+
    \frac{1}{2} p \left(g^{\alpha \beta} g^{\gamma \delta} - g^{\alpha \gamma} g^{\beta \delta} - g^{\alpha \delta} g^{\beta \gamma} \right) - \frac{1}{2} \Gamma_1 p q^{\alpha \beta} q^{\gamma \delta}
    \,.
\end{align}
Using Eq.~\eqref{eq:identity-for-sim} in Eq.~\eqref{eq:power-1} (Eq.~\eqref{eq:torque-1}) with $k_{\alpha\beta}= \mathcal{L}_{t}h_{\alpha\beta}$ ($k_{\alpha\beta}= \mathcal{L}_{\phi}h_{\alpha\beta}$) we obtain the final expressions for the power and torque densities
\begin{subequations}
\begin{align}
    \mathscr{P}
    &=
    -
    \frac{1}{2}
    \delta_{\xi}T^{\alpha\beta}
    \mathcal{L}_{t}h_{\alpha\beta}
    \label{eq:power}
    ,
    \\
    \mathscr{T}
    &=
    \frac{1}{2}
    \delta_{\xi}T^{\alpha\beta}
    \mathcal{L}_{\phi}h_{\alpha\beta}.
    \label{eq:torque-density}
\end{align}
\end{subequations}
We note that the angular-momentum source can also be written in a ``mechanical'' force balance form
\begin{align}
    \mathscr{T}_{\mathrm{mech}}
    =
    -
    F^{\mu}\mathcal{L}_{\phi}\xi_{\mu}.
\end{align}
However, for numerical diagnostics near and especially near corotation, the Eulerian form in Eq.~\eqref{eq:torque-density}
is preferable.
\subsection{Master equations for numerical implementation}\label{sec:master-eqs}
The Lagrangian perturbation theory described in the previous section is beneficial for mathematical analysis and to obtain balance laws for the spiral density waves.
For numerical implementations of the perturbed equations, it is useful to reduce the perturbations equations into a set of master equations. We now schematically sketch the derivation of these equations.

Let us first note that, we can ignore the continuity equation in our analysis below, since it completely decouples from the stress energy conservation. In presenting the master equations, we work in the Fourier domain in $(t,\phi)$, with Fourier transform conventions
\begin{subequations}
\begin{align}
    &f(t,r,\phi) = \sum_{m=-\infty}^{\infty} \hat{f}_m(\omega,r) e^{-i\omega t + i m \phi} \,, \\
    &\hat{f}_m(\omega,r) = \frac{1}{2\pi} \int_{0}^{2 \pi} f(t,r,\phi) e^{i \omega t - i m \phi} \, d\phi 
    \,,
\end{align}
\end{subequations}
where 
\begin{align}
    \omega = m \OmegaEMRI \,,
\end{align}
for an EMRI on a circular orbit [Eq.~\eqref{eq:OmegaEMRIKerr}].

We can obtain the master equations by using the following reduction scheme.
First solve for $\delta \hat{u}^{\phi}$ using the $\phi$ component of Euler equation. 
Eliminate $\delta \hat{u}^{\phi}$ from the $r$ component of the Euler and the energy conservation equation. This reduction results in a set of master equations for the variables $\delta \hat{U}$ and $ -i \delta  \hat{u}^r$ which can be schematically written as
\begin{align}\label{eq:master-equation-relativity}
    &\frac{d}{dr}\boldsymbol{y} = \boldsymbol{C}\cdot \boldsymbol{y} + \boldsymbol{S}
\end{align}
where $\delta \hat{U}$ is the Fourier transformed relativistic enthalpy perturbation of the fluid
\begin{align}
    \delta \hat{U} \equiv \frac{\delta \hat{p} }{\varepsilon + p} \,,
\end{align}
$\boldsymbol{C}$ is a (2 dimensional) matrix that depends on the background profiles, $\boldsymbol{S}$ is a (2 dimensional) gravitational source term due to the small object and
\begin{align}
    \boldsymbol{y} = \left(\delta \hat{U}, -i \delta  \hat{u}^r\right)^T \,.
\end{align}
The relativistic expressions for these matrices and sources are long, and so we provide them in the supplementary \texttt{Mathematica} notebook.
In the Newtonian limit, the matrices and sources are given by
\begin{subequations}
\begin{align}
    &\delta u_{\phi} = \frac{m \delta \hat{U}}{r \omega _d}+\frac{m \Phi _{\text{Newt}}}{r \omega _d}-\frac{i \delta \hat{u}^r \kappa ^2}{2 \omega _d \Omega } \,,\\
    &C_{0,0} = 
    \frac{2 m \Omega }{r \omega _d}+\frac{p'}{\Gamma_1  p}-\frac{\Sigma '}{\Sigma }\,,\\
    &C_{0,1} = -\omega _d+
    \frac{1}{\omega _d}
    \bigg[
    \kappa ^2+\frac{p' \left(-\frac{\Sigma  p'}{\Gamma_1  p}+\Sigma '\right)}{\Sigma ^2}
    \bigg]
    \,,\\
    &C_{1,0} = -\frac{m^2}{r^2 \omega _d}+\frac{\omega _d \Sigma }{\Gamma_1  p}\,,\\
    &C_{1,1} = -\frac{1}{r}-\frac{m \kappa ^2}{2 r \omega _d \Omega }-\frac{p'}{\Gamma_1  p}\,,\\
    &S_0 = \frac{2 m \hat{\Phi}_{\text{Newt}} \Omega }{r \omega _d}-\hat{\Phi}_{\text{Newt}}' \,,\\
    &S_1 = -\frac{m^2 \hat{\Phi}_{\text{Newt}}}{r^2 \omega _d} \,,
\end{align}
\end{subequations}
where $\kappa$ is the epicyclic frequency, 
\begin{align}
    \omega_{d} \equiv \omega - m \Omega
\end{align}
is the Doppler shifted forcing frequency as seen in a frame comoving with a fluid parcel, $\Sigma$ is the surface density and $\hat{\Phi}_{\mathrm{Newt}}$ is the Fourier-transformed Newtonian gravitational potential of the EMRI.
Observe that the Newtonian equations are singular at a corotation resonance $\omega_{d} = 0$. The corotation resonances are true singularities of the perturbed equations and require special care.
The exact singular structure seen in the Newtonian equations also persists in the relativistic regime and we describe how to treat them in Sec.~\ref{sec:numerics} below. 
\subsection{Relativistic gravitational potential of the small object}\label{sec:EMRI-potential}
The gravitational potential generated near a moving object in general relativity can be separated into a Newtonian-like singular potential that diverges as we approach the location of the object and a radiative piece that describes freely propagating waves generated by the object.
In this paper, we restrict our attention to understanding the effect of the singular Newtonian-like potential of the object on density waves in an accretion disk.
Analytic expressions for the singular field near the worldline of an object are available from~\cite{Heffernan_2012,Pound_2014}. 
A review of the application of these techniques in the accretion-disk EMRI interactions case is available in Sec. IV of~\cite{HegadeKR:2025dur}.

Suppose that the small object is moving on a worldline $w^{\mu}(t)$ in Kerr spacetime with four velocity $v^{\mu}(t)$.
The singular Newtonian-like gravitational potential generated the small object at location $x^{\mu}$ to leading order is given by
\begin{align}\label{eq:sing-func-expr-1}
    h_{\mu\nu}(x)
    &\approx
    \frac{2 \, \mEMRI}{\textsf{s}_0} \left(g_{\mu \nu}(w) + 2 v_{\mu} v_{\nu} \right)
\end{align}
where $\mEMRI$ is the mass of the secondary and $\textsf{s}_0$ is approximately the coordinate distance between the location $x$ and the worldline $w^{\mu}$. 
There is considerable freedom in defining $\textsf{s}_0$ and one usually obtains very high order Taylor series expansions in infinitesimal coordinate distances from the worldline to obtain accurate expressions~\cite{Wardell_2012,Pound_2014,Bourg_2024}.

In this work, we use a simple approximation to $\textsf{s}_0$ for circular geodesics in Kerr spacetime. The worldline $w^{\mu}(t)$ and four velocity $v^{\mu}(t)$ of an equatorial circular geodesic in $(t,r,\phi,z)$ coordinates is given by~\cite{Bourg_2024}
\begin{align}
    &w^{\mu}(t) = \left(t, \aEMRI,  \OmegaEMRI t, 0 \right)\,,\\
    &v^{\mu}(t) = \left(\frac{dt}{d\tau},0,\OmegaEMRI \frac{dt}{d\tau},0 \right) \,,\\
    &\frac{dt}{d\tau} = \frac{a_{\star}^{3/2}+\chi }{a_{\star}^{3/4} \sqrt{\sqrt{a_{\star}} (a_{\star}-3)+2 \chi }} 
    \,,\quad 
    a_{\star} \equiv \aEMRI/M \,.
\end{align}
To approximate $\textsf{s}_0$, we first introduce Cartesian coordinates close to the equator
\begin{align}
    y^{\mu} = \left(t, r \cos(\phi), r \sin(\phi), z \right)
    \,,
\end{align}
and obtain the infinitesimal coordinate distance from the circular geodesic worldline in the equatorial plane 
\begin{align}
    &\Delta y^{\mu} \equiv \nonumber\\
    &\bigg(0, r \cos(\varphi) - \aEMRI \cos(\OmegaEMRI t), r \sin(\varphi) - \aEMRI \sin(\OmegaEMRI t),0 \bigg)\,.
\end{align}
Our approximation to $\textsf{s}_0$ is given by
\begin{align}
    \textsf{s}_0^2 \approx \left.\left[g_{\mu \nu} + v_{\mu} v_{\nu} \right] \frac{\partial x^{\mu}}{\partial y^{\alpha}} \frac{\partial x^{\nu}}{\partial y^{\beta}} \right|_{w}
    \Delta y^{\alpha} \Delta y^{\beta}
    +
    \epssoft^2
    \,,
\end{align}
where $x^{\mu} = (t,r,\phi,z)$ and $\epssoft$ is a Plummer softening length.
We use a simple prescription for the softening length
\begin{align}\label{eq:softening-prescription}
    \epssoft = \aEMRI h_0 \bsoft\,,
\end{align}
where $\bsoft$ is a dimensionless number typically in the range $0.1-0.7$.
The explicit expression for $\textsf{s}_0^2$ for points in the equatorial plane are given in Appendix~\ref{appendix:ssq-expr}.\footnote{Qualitatively, effective softening of the potential in 2D calculations can be physically meaningful because the column-integrated force on a disk at fixed cylindrical radius will be less than the product of the surface density and the force evaluated at the disk midplane. However, the resulting functional form would still differ from that introduced here and softening would still be required to avoid singularities near point mass.} 

Using them, one can evaluate the Fourier transformed potentials required for the perturbation equations as
\begin{align}
    \hat{h}^{\mu \nu}_{m}
    =
   \frac{1}{2 \pi}
    \int_{0}^{2\pi}
    h^{\mu \nu} \cos(m \Psi) d \Psi
    \,.
\end{align}
\subsection{Numerical scheme, boundary conditions and diagnostics}\label{sec:numerics}
In this section, we describe the numerical technique used to integrate the master equations derived in Eq.~\eqref{eq:master-equation-relativity}.
As we mentioned above, the master equations have a non-integrable singularity at the corotation location $r_c$
\begin{align}
    r_c : \omega_{d}(r=r_c) = 0 \,. 
\end{align}
To render the relativistic equations integrable at $r_c$ we replace every occurrence using a Landau prescription \cite{K&P_1993,2024ApJ...968...28T}
\begin{align}
    \omega_{d} \to \omega_{d} + i \epsilon(r)
    \,,
\end{align}
where $\epsilon(r)$ is a Gaussian function centered around the corotation location $r_c$ with amplitude $\epsilon_0$ and width $\sigma_{\epsilon}$
\begin{align}
    \epsilon(r)
    =
    \epsilon_p \Omega_\bullet(r_c)
    \exp\left[
        -\frac{(r-r_c)^2}{2\sigma_\epsilon^2}
    \right],
\end{align}
Typically, in our implementation, we set $\epsilon_p = 10^{-5}$ and $\sigma_{\epsilon} = 0.1 \aEMRI$.

Even with the Landau prescription, the corotation singularity is very strong and this requires a specialized strategy to integrate the master equations. We have tested different methods such as direct shooting method, multiple shooting methods and implicit numerical integration.
We find that a second order implicit integration scheme with a specialized radial grid is the most robust and stable approach.

The scheme starts by generating a specialized grid in $r$ that clusters around corotation. We describe the grid generation method in detail in Appendix~\ref{appendix:grid}.
Let us denote the grid points using index $i$.
Integrate the master equations in the interval $[r_{i}, r_{i+1}]$, and use the mid-point rule to obtain
\begin{align}
    &\frac{\boldsymbol{y}_{i+1}-\boldsymbol{y}_{i}}{\Delta r_i}
    =
    \boldsymbol{C}_{i+1/2} \cdot \frac{\boldsymbol{y}_{i} + \boldsymbol{y}_{i+1}}{2}
    +
    \boldsymbol{S}_{i+1/2}
    \,,
    \nonumber\\
    \Leftrightarrow
    &
    \left(\mathbb{I}_{2\times2} -  \frac{\Delta r_i}{2} \boldsymbol{C}_{i+1/2} \right) \cdot\boldsymbol{y}_{i+1}
    \nonumber\\
    &-
    \left(\mathbb{I}_{2\times2} + \frac{\Delta r_i}{2} \boldsymbol{C}_{i+1/2} \right) \cdot\boldsymbol{y}_{i}
    =
    \Delta r_i \boldsymbol{S}_{i+1/2}
    \,,
    \label{eq:discrete-master-equation}
\end{align}

where $\mathbb{I}$ is the identity matrix and
\begin{align}
    \Delta r_{i} = r_{i+1} - r_{i} \,.
\end{align}
To close the discretized equations, we use the WKB wavenumber to choose waves outgoing from the orbit of the small object as the boundary conditions at the edge of the domain.
Specifically, we first calculate the eigenvalues $k$ and eigenvectors $\boldsymbol{Y}$ of $\boldsymbol{C}$ at the edge of the domain
\begin{align}
    \boldsymbol{C} \cdot \boldsymbol{Y} = k \boldsymbol{Y}
    \,,
\end{align}
and choose the eigenvector for which $\mathrm{Im}\left[k\right] <0$. This eigenvector serves as the outgoing solution and we set set
\begin{align}
    \boldsymbol{y} = c_1 \boldsymbol{Y} \,.
\end{align}
Numerically, this condition can be implemented as a Neumann-type boundary condition by finding the vector perpendicular to $\boldsymbol{Y}$ and setting
\begin{align}
    \boldsymbol{y} \cdot \boldsymbol{Y}_{\perp} = 0\,,
\end{align}
at the edge of the domain.

For diagnostic purposes, we track the advected angular momentum flux and the torque. One useful identity which simplifies the calculation enormously is 
\begin{align}\label{eq:jrphi-identity}
    j^{r}_{J} = \delta p \delta u^{r} u_{\phi} + \left(\varepsilon + p \right) \delta u^{r} \delta u_{\phi}
    \,.
\end{align}
Physically, this states that the Noether current obtained from Lagrangian perturbation theory is equal to the advected angular momentum obtained from the variation of the stress tensor.
One can show that a similar result holds for the energy flux
\begin{align}
    j^{r}_{E} = -\delta p \delta u^{r} u_{t} - \left(\varepsilon + p \right) \delta u^{r} \delta u_{t}
    \,.
\end{align}

Let us denote the azimuthally averaged radial AMF by $F_J$ and the radially integrated torque by $\mathfrak{T}(r)$.
For circular orbits, we have
\begin{align}\label{eq:AMF-Torque-Balance}
    \frac{d}{dr} F_J = \frac{d \mathfrak{T}}{dr} \,.
\end{align}

The explicit expressions for $F_J$ and $\mathfrak{T}$ can be obtained by using Eqs.~\eqref{eq:angular-momentum-current}, \eqref{eq:torque-density} and \eqref{eq:jrphi-identity} 
\begin{subequations}
\begin{align}
    &F_J = \left< \sqrt{-g} j^{r}_{\phi} \right>
    =
    \sqrt{-g}
    \bigg[
    u_{\phi}
    \left< 
    \delta p \delta u^{r}\right>  + \left(\varepsilon + p \right) \left<\delta u^{r} \delta u_{\phi}
    \right>
    \bigg]
    \,,\\
    &\mathfrak{T}(r) = \int_{r=a}^{r} dr' \sqrt{-g} \left<\mathscr{T} \right>
    \nonumber\\
    &=
    \frac{1}{2}\int_{r=a}^{r} dr'  \sqrt{-g} \left< \delta_{\xi} T^{\alpha \beta} \partial_{\phi} h_{\alpha \beta} \right>
    \,.
\end{align}
\end{subequations}
where $\left<\cdot\right>$ denotes the integral over $\phi$.
To compute the average, one can use standard Fourier identities to show that for real $X$, $Y$
\begin{align}
    \left< X Y \right> \equiv \int_{0}^{2\pi} X Y d\phi = 2 \pi \hat{X}_0 \hat{Y}_0 + 4 \pi \sum_{m=1}^{\infty} \mathrm{Re} \left[ \hat{X}_m \hat{Y}^{*}_{m} \right] \,.
\end{align}
The total torque on the grid can be computed as
\begin{align}
    T  = \mathfrak{T}(r_{\mathrm{out}}) - \mathfrak{T}(r_{\mathrm{in}}) \,. 
\end{align}
It is useful to separate the total torque into contributions from Lindblad resonances and the corotation torque. This decomposition can be performed exactly for thin pressureless disks~\cite{HegadeKR:2025dur,HegadeKR:2025rpr,Duque:2025yfm}. For fluid disks studied here we must rely on an approximate decomposition which relies on the AMF and torque jump near corotation~\cite{FR-2025}.
Let us denote the jump in the torque density and AMF near corotation by $\Delta_{\mathrm{c}} T$ and $\Delta_{\mathrm{c}} F_J$.
The torque due to Lindblad resonance is defined as
\begin{align}
    T_{L}
    &=
    T
    -
    \left(
    \Delta_{\mathrm{c}} T
    -
    \Delta_{\mathrm{c}} F_J
    \right)
    =
    F_J(r_{\mathrm{out}})
    -
    F_{J}(r_{\mathrm{in}})
    \,.
\end{align}
The second equality is a consequence of Eq.~\eqref{eq:AMF-Torque-Balance}.
\begin{figure*}
    \centering
    \includegraphics[width=0.99\linewidth]{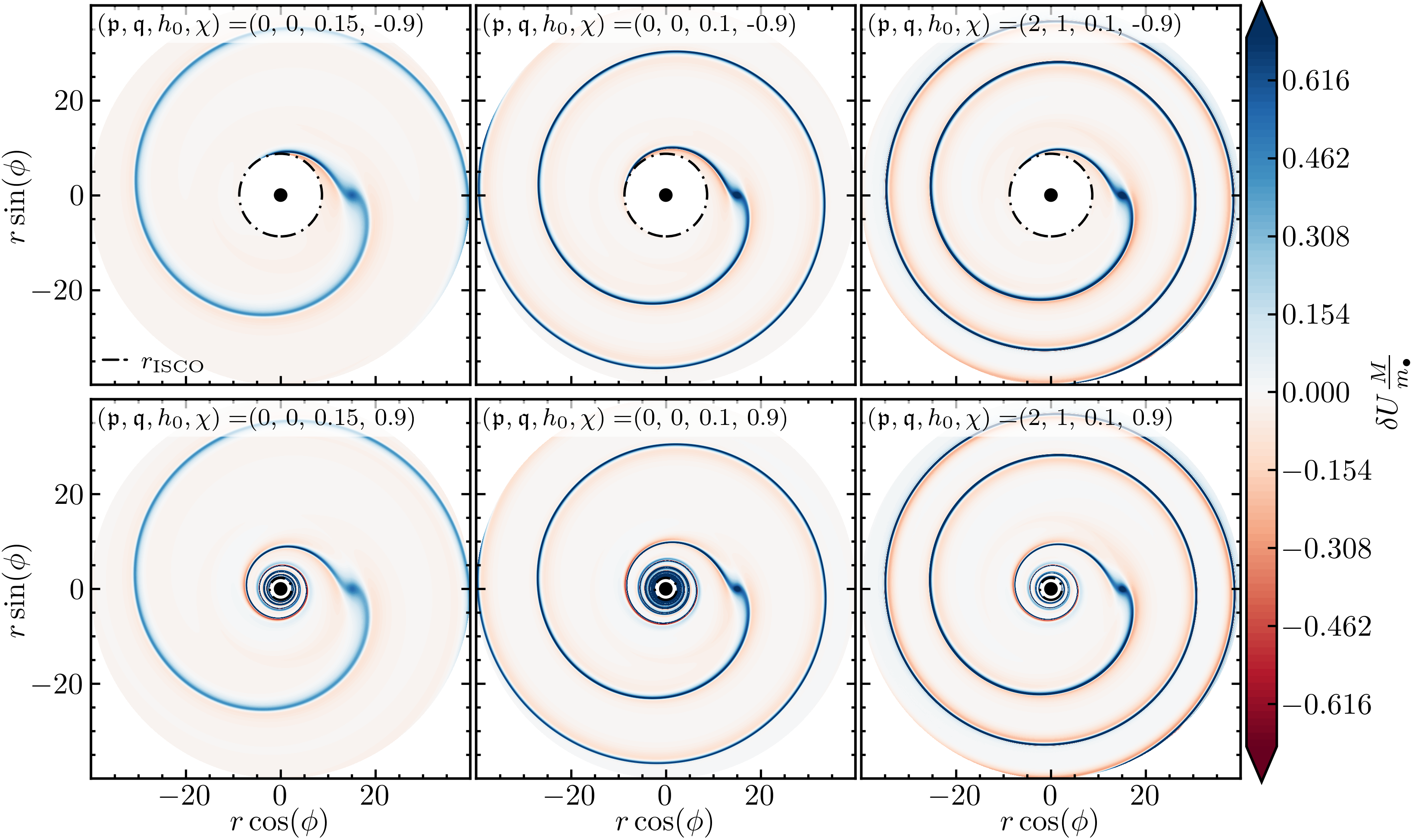}
    \caption{The linear response of various accretion disks to an EMRI at $\aEMRI= 15 M$, visualized via the resulting enthalpy perturbation. The spiral arms are truncated at the ISCO (black dash-dot line), and wind more tightly in thinner disks. In disks with $\partial_rc_s<0$, the inner spiral arm winds more loosely, while the outer spiral arm winds more tightly.}
    \label{fig:spirals}
\end{figure*}

\begin{figure*}
    \centering
    \includegraphics[width=0.99\linewidth]{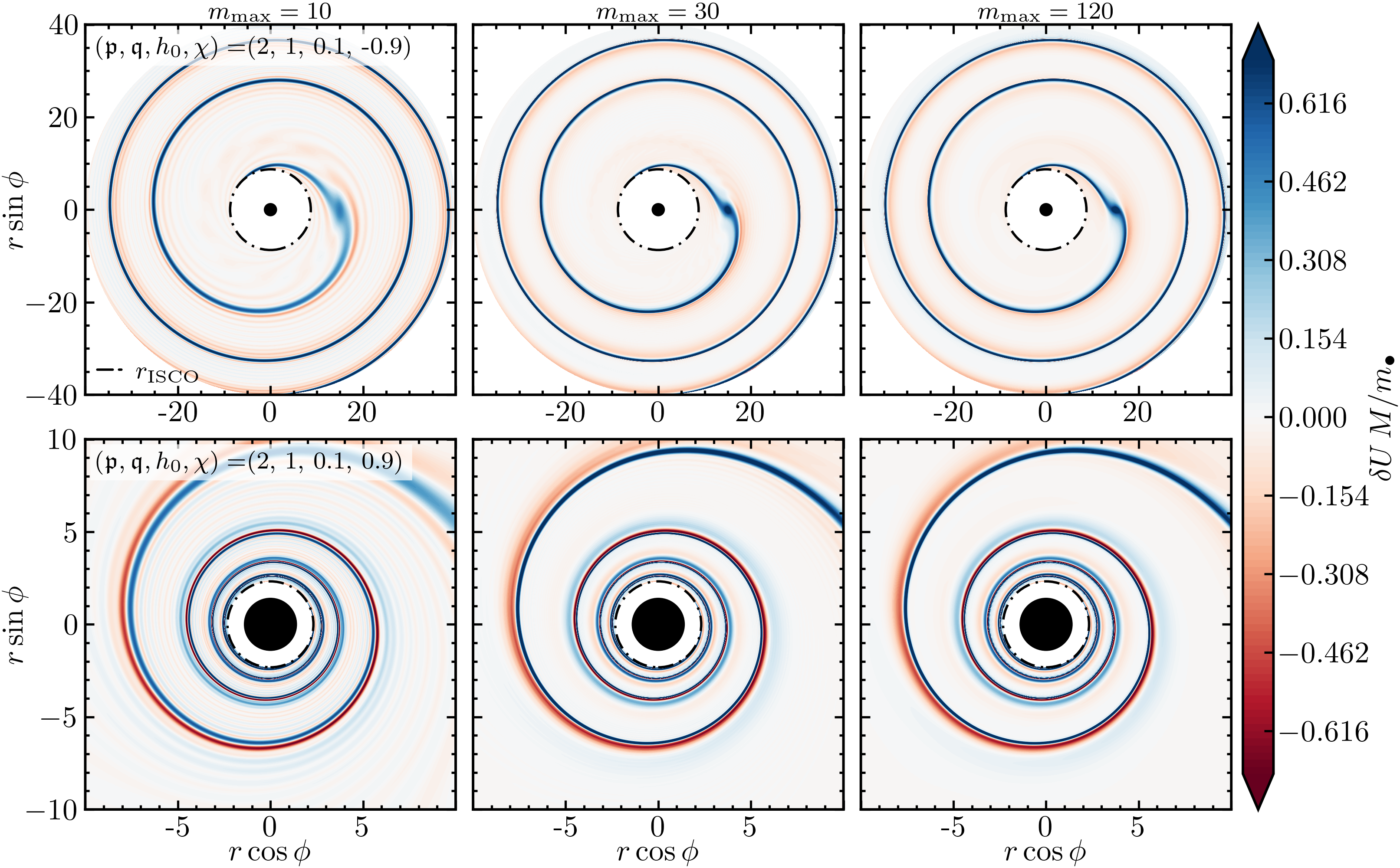}
    \caption{The convergence of the spiral patterns in a $\pdisk=2,\qdisk=1,h_0=0.1$ disk with respect to the maximum azimuthal mode number $m$ for an EMRI located at $a_{\bullet} = 15 M$. The top (bottom) panel shows the spiral waves for $\chi = -0.9$ ($\chi = 0.9$). Fourier ringing and blurring of the co-orbital region are obvious for in the first two columns. In the $\chi=0.9$ case, in which the spiral can propagate inward for multiple wavelengths before being truncated by the ISCO, the single spiral arm decoheres into secondary and tertiary spirals, as seen by the spiral splitting into multiple peaks and troughs in the enthalpy perturbation in the bottom-right panel. However, when fewer modes are included, it can be difficult to distinguish these genuine features from Fourier ringing in the synthesized disk response. }
    \label{fig:spirals_conv}
\end{figure*}

\section{Results}\label{sec:results}
In this section, we will present exemplary calculations of linear disk-EMRI interactions, both in terms of the morphological response of the disk to the perturber and of the associated angular momentum exchange between the two, followed by some illustrative parameter surveys and comparisons with analytical work. 

Let us first define the characteristic flux $F_{J,0}$ and torque density $F_{J,1}$ values~\cite{GT-disc-satellite-interaction}, which will be used to quote the results in this paper
\begin{subequations}
\begin{align}
    &F_{J,0} = \frac{\varepsilon_0 \aEMRI^4 \OmegaEMRI(\aEMRI)^2}{h_0^3} q^2 \,,\\
    &F_{J,1} = \frac{F_{J,0}}{\aEMRI} = \frac{\varepsilon_0 \aEMRI^3 \OmegaEMRI(\aEMRI)^2}{h_0^3} q^2 
    \,.
\end{align}
\end{subequations}
Note the dependence of these quantities on $\varepsilon_0\equiv\varepsilon(a_\bullet)$, and $a_\bullet$ generally; while the results quoted below scale out these units to focus on the dimensionless parameters that shape disk-EMRI interactions --- $a_\bullet$, $\chi$, $h_0$, $\mathfrak{p}$, and $\mathfrak{q}$ --- these quantities will vary during realistic EMRIs, and should be kept in mind when interpreting the results presented below.

As described in Sections \ref{sec:EMRI-potential} and \ref{sec:numerics}, the calculations herein depend on the numerical parameters $b_{\rm soft}$ (the softening parameter from Eq.~\eqref{eq:softening-prescription}) and the pole displacement $\epsilon_p$. In the calculations below we fix these parameters to $\bsoft=0.6$ and $\epsilon_p=10^{-5}$. This value for $\epsilon_p$ is chosen largely for convenience, since smaller values would require a finer grid spacing near corotation, and as long as $\epsilon_p\ll1$ its effect on the solution is negligible. The value chosen for $b$ has a more significant effect, since a smaller value will increase the gravitational potential of the secondary and thus the torque, particularly for higher-wavenumber modes that resonate closest to the perturber. Our nominal value of $\bsoft=0.6$ was chosen because in 2D Newtonian calculations it closely matches the total torques measured in analogous three-dimensional calculations \citep[e.g.,][]{Cordwell_2025}. A discussion of the dependence of the torque on the $\bsoft$ and $\epsilon_p$ is presented in Appendix~\ref{appendix:convergence}. 

The remainder of this section is organized as follows. We first present our exemplary calculations in Sec.~\ref{sec:exemplary-calc}. Next, we survey the impact of relativistic effects in Sec.~\ref{sec:surveys}. Finally, we compare our results to pressureless models in Sec.~\ref{sec:comparison}.

\subsection{Exemplary Calculations}\label{sec:exemplary-calc}

In Fig.~\ref{fig:spirals} we illustrate how the morphology of the spiral arms excited in disks by EMRIs depends on the disk parameters and SMBH spin. Fundamentally, the linear disk response manifests via sound waves which are excited at and propagate away from Lindblad resonances at the local sound speed; the winding of spiral arms through the disk is thus governed by the characteristic local aspect ratio $h_0$ and its power-law index $\mathfrak{q}$, lower-pressure fluid leading to more tightly wound spirals. The most prominent effect of the SMBH spin is truncating the inner disk at the $\chi$-dependent ISCO, although it also results in more subtle changes to the disk response insofar as SMBH spin affects the angular and radial epicyclic frequencies of the disk and thus the resonant locations for various modes. 

Figure \ref{fig:spirals_conv} illustrates how the single-armed spiral excited by the EMRI arises as the constructive interference of hundreds of modes. 
For the disks considered in these figures, most of the large-scale Fourier ringing subsides by $m\geq40$, though some subtle ringing can persist in the outer and inner regions even to $m\gtrsim100$. 
In the inner regions of these disks the spiral arm can decohere, splitting into secondary and even tertiary spirals, as also seen in the Newtonian case \cite{Miranda_2019}. 
For example, the spiral arm is launched near the secondary as a single arm (see the positive-negative pattern near the secondary in Figure \ref{fig:spirals}); however, after propagating inward, the spiral splits and displays a pattern of multiple peaks and troughs (as seen cleanly in the bottom-right panel of Figure \ref{fig:spirals_conv}). 
When only a few modes are considered, the secondary and tertiary spirals can be comparable in prominence to Fourier ringing, but after summing hundreds of modes, they are clearly robust. 
Although it is necessary to compute hundreds of modes to model the disk response with high fidelity, the orbits of quasi-circular EMRIs are only affected by the reflection-asymmetric response of the disk about the EMRI's orbit; as we show below, accurate predictions of orbital evolution can be achieved by calculating far fewer modes.

\begin{figure}
    \centering
    \includegraphics[width=0.99\linewidth]{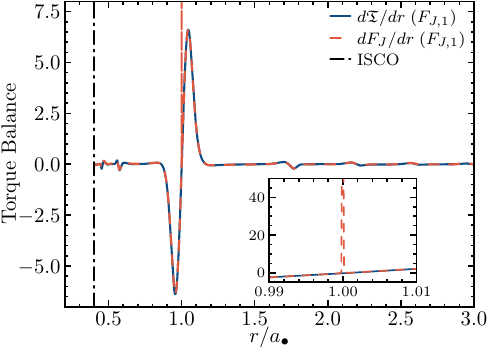}
    \caption{Angular momentum balance in disk-EMRI interactions. The red-dashed curves are used to denote the AMF density and the blue solid curves are the torque density. Observe that they are perfectly on top of each other except near corotation $r \sim \aEMRI$ where angular momentum is deposited. The inset shows the spike in the AMF density near corotation. }
    \label{fig:angular_momentum_balance}
\end{figure}

In Fig.~\ref{fig:angular_momentum_balance} we show how the EMRI exchanges angular momentum with the disk around a Schwarzschild black hole. The figure illustrates the balance between the AMF through the disk and the torque on the disk due to the EMRI for $(\pdisk, \qdisk) = (0,0), \chi = 0, h_0 = 0.05$ and $\aEMRI = 15 M$. Observe that $d F_J /dr $ (red dashed curve) lies perfectly atop $d \mathfrak{T}/dr$ (solid blue curve) everywhere away from $r\sim \aEMRI$, demonstrating the consistency of our angular momentum balance formulation. 
At corotation ($r \sim \aEMRI$), there is a huge jump in $d F_J /dr $ (see, the zoomed inset) which indicates the deposition of angular momentum; the sharpness of this near-discontinuity at corotation is acutely sensitive to gradients in the background flow. Although treating this singularity at corotation is analytically tractable in some special cases \cite[e.g.,][]{Dyson:2026ddd}, our general numerical treatment resolves this sharp feature using targeted mesh refinement and a Landau pole displacement prescription, the accuracy and robustness of which are tested in Appendix \ref{appendix:grid}. 

\subsection{Surveying Relativistic Effects}\label{sec:surveys}
Having examined the preceding examples of disk-EMRI interactions, we now turn to a systematic investigation into the importance of various relativistic effects. While SMBH spin obviously affects the location of the inner edge of the disk, $\chi$ also impacts the angular velocity profile of the fluid, and relativistic corrections to Newtonian gravity generically detune the angular frequency and radial epicyclic frequency of the disk, shifting the locations of Lindblad resonances \cite[e.g.,][]{HegadeKR:2025dur,HegadeKR:2025rpr}. 

\begin{figure}
    \centering
    \includegraphics[width=0.99\linewidth]{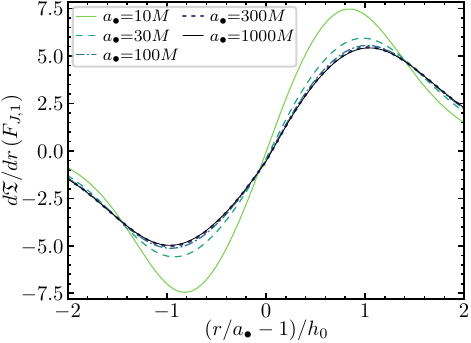}
    \caption{
    Torque density profiles near corotation for various $\aEMRI/M$ for $\chi=\mathfrak{p}=\mathfrak{q}=0$. In the nearly Newtonian limit ($\aEMRI=1000M$) the peaks of the torque density are offset by about $\pm h_0\aEMRI$ from corotation; in the strongly relativistic case ($\aEMRI=10M$), the peaks of the torque density shift inward by $\sim \aEMRI h_0/5$ and increase in magnitude by a factor of $\sim1.5$.
    }
    \label{fig:dtdr_ap}
\end{figure}

We first examine relativistic effects in the Schwarzschild limit in Figure.~\ref{fig:dtdr_ap}, which illustrates how the torque density near the EMRI shifts between the essentially Newtonian limit ($a_\bullet = 1000\,M$) and the strongly relativistic regime ($a_\bullet = 10\,M$) for a thin disk with parameters $\pdisk=\qdisk = 0$ and characteristic scale height $h_0 = 0.05$. As resonances shift relatively closer to the perturber due to relativistic changes in the fundamental frequencies of the background disk, the gravitational forcing becomes more localized near the EMRI and the disk response becomes stronger. However, since the net torque results from a near-cancellation of negative contributions from the outer disk and positive contributions from the inner disk, the increased amplitude of the torque density profile in the relativistic regime displayed in Fig.~\ref{fig:dtdr_ap} does not necessarily lead to stronger effects on EMRI migration. 

\begin{figure}
    \centering
    \includegraphics[width=0.99\linewidth]{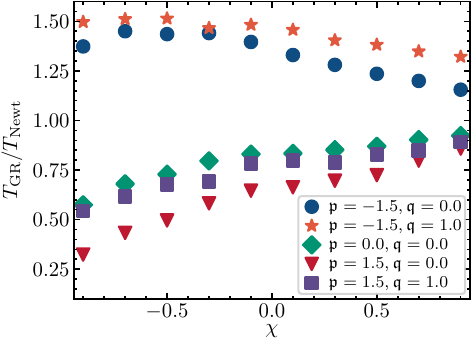}
    \caption{Ratio of relativistic torque to the Newtonian torque for disk profiles for a scale height of $h_0 = 0.05$ and $\aEMRI = 12 M$ as a function of the spin of the SMBH. Observe that disks with negative (positive) surface density slope $\pdisk>0$ $(\pdisk<0)$ can decrease (increase) the relativistic torque compared to the Newtonian calculation.}
    \label{fig:TGR_vs_TNewt}
\end{figure}

We illustrate the difference between the torques in the relativistic and Newtonian prescription as a function of spin for an EMRI located at $\aEMRI = 12 M$ in Fig.~\ref{fig:TGR_vs_TNewt}. The disk density and sound speed profiles in this figure vary but we have fixed the characteristic scale height to $h_0 = 0.05$.
Note that for disks with radially increasing energy density profiles ($\mathfrak{p}<0$) the overall torque can increase in the relativistic regime (as the slightly dominant outer resonances become even stronger), while for disks with neutral or decreasing energy density profiles we find that net torques tend to become weaker. For sufficiently extreme density profiles the disk torque may cross zero and eventually change sign, as predicted by earlier analytical estimates based on dust disks \cite{HegadeKR:2025rpr,HegadeKR:2025dur,Duque:2025yfm}; however, we find no such torque reversal, at least for disk thicknesses $0.05\leq h_0\leq 0.15$, and discuss this discrepancy in more detail in Sec.~\ref{sec:comparison}.

\begin{figure}
    \centering
    \includegraphics[width=0.99\linewidth]{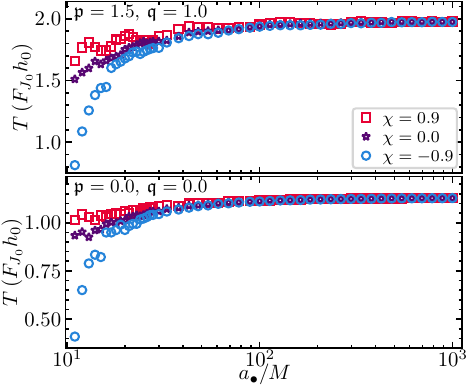}
    \caption{Variation in the total torque as a function of $\aEMRI$ for different values of $\chi$ for a disk of scale height $h_0 = 0.05$. Relativistic effects are important as we approach $\aEMRI \sim \mathcal{O}(50 M)$; the retrograde orbits $(\chi<0)$ face stronger relativistic effects since the ISCO is closer to the EMRI orbit compared to the prograde $(\chi>0)$ and the non-spinning $(\chi=0)$ cases.}
    \label{fig:torque_chi_ap}
\end{figure}

Qualitatively, the effect of SMBH spin should become most apparent very near the SMBH, since spin-dependent metric terms fall off relatively quickly beyond a few tens of $M$, while deviations between a flat and Schwarzschild background can appear at larger distances, closer to $\sim100\,M$. We display the dependence of the net torque on the disk on $\chi$ and $\aEMRI$ in Fig.~\ref{fig:torque_chi_ap} for a disk of characteristic scale height $h_0 = 0.05$; while this exemplifies the general trend described earlier, spin can have some subtle effects even for larger orbital separations for some disk profiles (e.g., the slight oscillations of the total torque at large separations for $\chi=0.9, \mathfrak{p}=1.5,\mathfrak{q}=1.0$ relative to $\chi=0$ and $\chi=-0.9$ for the same disk parameters).
Because the lowest-$m$ modes excited by the EMRI have very long wavelengths, these low-m modes can be affected by subtle differences in the inner disk even when $a_\bullet \gg M$, and yet more so in disks with $\mathfrak{p}>0$ such that inner spirals wind even less tightly. Still, at $\aEMRI \lesssim 20 M$ SMBH spin shifts resonances within the disk enough to cause order-unity changes in the net angular momentum exchanged between the EMRI and disk. 

\begin{figure}
    \centering
    \includegraphics[width=0.99\linewidth]{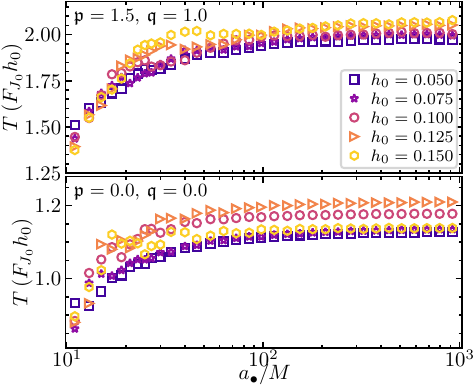}
    \caption{Variation in the torque with $h_0$ and $a_\bullet$ for disks around a SMBH of spin $\chi = 0$. Thicker disks are impacted strongly by the inner regions of the disk leading to some oscillatory behavior. Observe that the torque monotonically increases until $h_0 = 0.125$ and then decreases for very thick disks $h_0 = 0.15$.}
    \label{fig:torque_ap_h0}
\end{figure}

We display the results of a similar survey in Fig. \ref{fig:torque_ap_h0}, which focuses on a single SMBH spin $\chi = 0$ but varies the characteristic disk scale height $h_0$. While the torques on thicker disks show some oscillatory variation with $a_\bullet/M$ (as seen around $\aEMRI \sim 20 M)$, as low-m modes in such disks probe the strongly non-scale-free inner regions of the disk, torques on thinner disks tend to vary smoothly with $a_\bullet/M$. 
\subsection{Comparison with pressureless models}\label{sec:comparison}

Here, we present a comparison of our linear theory with the analytical and numerical calculations of relativistic disk-EMRI interactions that preceded it~\cite{HegadeKR:2025dur,HegadeKR:2025rpr,Duque:2025yfm}.
In this comparison, it is important to note that~\cite{HegadeKR:2025dur,HegadeKR:2025rpr,Duque:2025yfm} assumed that the i) accretion disks were pressureless, ii) calculated only the Lindblad torque, and iii) assumed a simplified torque cutoff parameter when calculating the total torque. Pushing our calculations to the nearly pressureless regime requires a very small disk scale height and is computationally challenging. We include a few calculations here with scale heights as low as $h_0=0.03$, but focus mostly on slightly thicker disks with $h_0=0.05$ and $h_0=0.15$. 

\begin{figure}
    \centering
    \includegraphics[width=0.99\linewidth]{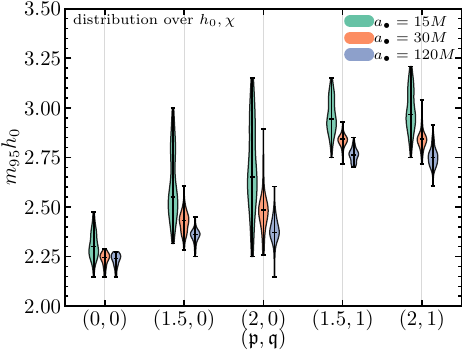}
    \caption{Mode number at which $95\%$ of the total cumulative torque is attained for different values of $(\mathfrak{p},\mathfrak{q})$. The distribution over $h_0$ and $\chi$ is obtained by scanning $h_0 \in [0.05, 0.15]$ and $\chi \in [-0.9,0.0, 0.9]$. Observe that $m_{95} h_0$ is approximately constant and the largest variation is observed when the EMRI is closest to the SMBH.}
    \label{fig:m_95_plot}
\end{figure}

\begin{figure}
    \centering
    \includegraphics[width=0.99\linewidth]{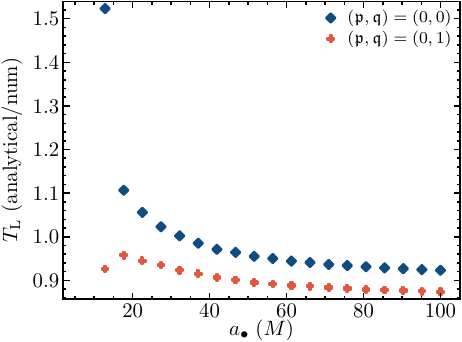}
    \caption{Comparison between the Lindblad torque obtained from linear theory and analytical results from~\cite{HegadeKR:2025dur,HegadeKR:2025rpr} in Schwarzschild spacetime for $h_0 = 0.05$. The analytical torque cutoff parameter has been set to $m_{\mathrm{max}} = m_{45}$. The difference between the analytical results and the numerical result increases as we approach the ISCO.}
    \label{fig:num_vs_analytical}
\end{figure}

First, let us discuss the issue of the torque cutoff parameter. The asymptotic formula presented in~\cite{HegadeKR:2025dur,HegadeKR:2025rpr} requires a truncation at a large but finite value of $m$ to prevent the result diverging.
Schematically, the analytical per-mode Lindblad torque scales as
\begin{align}
    T_{L,m}^{\mathrm{analytical}} \propto m \,.
\end{align}
The total analytical Lindblad torque is estimated as
\begin{align}
    T_{L}^{\mathrm{analytical}} = \sum_{m=1}^{m_{\mathrm{max}}} T_{L,m}^{\mathrm{analytical}}
    \propto m_{\mathrm{max}}^2 
    \,,
\end{align}
where we have used the asymptotic formula $\sum_{m=1}^{m_{\mathrm{max}}} m~\!\!\!\!\!\!\!=~\!\!\!\!\!\!\!m_{\mathrm{max}}^2/2 + \mathcal{O}(m_{\mathrm{max}})$; see the discussion near Eq. (24) of~\cite{HegadeKR:2025rpr}.
The exact analytical result in Kerr spacetime can be obtained from Eqs. (18 b) and Eq. (20) of~\cite{HegadeKR:2025rpr}.

This cutoff value $m_{\mathrm{max}}$ is called the torque cutoff parameter, and its value was set to\footnote{Note that~\cite{HegadeKR:2025dur,HegadeKR:2025rpr} use $j_{\mathrm{max}}$ to denote their torque cutoff parameter instead of $m_{\mathrm{max}}$.}
\begin{align}
    m_{\mathrm{max}} \approx \frac{1}{h_0} \,.
\end{align}
To compare our linear theory to analytical calculations, we need to estimate a torque cutoff parameter which can be used in the analytical calculations. 
We define $m_K$ as the smallest azimuthal mode number for which the cumulative torque reaches $K\%$ of the final torque obtained after summing all computed modes.
In Fig.~\ref{fig:m_95_plot} we present the results for the value of $m_{\mathrm{95}}$ for a few different disk surface density and sound speed slopes. For each $(\pdisk, \qdisk)$ we survey for $h_0 \in \left[0.05, 0.15\right]$, $\chi \in \left\{-0.9, 0.0, 0.9 \right\}$ and $\aEMRI \in \left\{15, 30, 120 \right\} M$. Observe that $\mninefive$ scales as $1/h_0$ and the scaling is approximately universal across different values of spins and $\aEMRI$ at a given value of $(\pdisk, \qdisk)$. As the EMRI location gets closer to the ISCO, relativistic effects impact the cutoff parameter, and we see a larger variation in $\mninefive$. Note that $m_{95}$ also depends on the gravitational softening, smaller values of $b$ requiring larger values of $m$ to achieve convergence; we illustrate this dependence in Appendix~\ref{appendix:convergence}.

In Fig.~\ref{fig:num_vs_analytical} we compare the analytical torque to our numerical results for disk parameters $h_0 = 0.05$, $(\pdisk, \qdisk) = (0,0)$ and $(\pdisk, \qdisk) = (0,1)$ as a function of $\aEMRI$ in Schwarzschild spacetime.
Unlike the numerical results, the analytical models incorporate no notion of the disk pressure or its gradient, and they depend on $q$ only implicitly through the choice of $m_{\mathrm{max}}$.
To facilitate a close match, with $\pdisk = \qdisk = 0$ we picked 
\begin{align}
    m_{\mathrm{max}} = m_{45} \sim \frac{0.35}{h_0 
    }\,.
\end{align}
Even after tuning $m_{\mathrm{max}}=m_{45}$, the match between the analytical and numerical results is still not perfect, and the mismatch increases when the EMRI is close to the ISCO or when there are sound speed gradients in the disk (orange $+$ markers). 
The increase in the mismatch as we approach the ISCO is expected. Our treatment of the inner boundary condition near the ISCO is approximate, and lower m-modes are especially susceptible to changes in the inner-boundary condition. The analytical results, on the other hand, assume that the waves are in the WKB approximation everywhere.
These differences suggest that the analytical formula, while useful to gain intuition, needs to incorporate pressure effects, and potentially other physics,
before it can be used in data analysis pipelines~\cite{Fantoccoli:2026idl}. 

\begin{figure}
    \centering
    \includegraphics[width=0.99\linewidth]{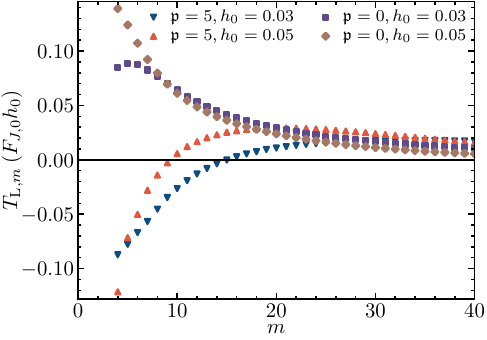}
    \caption{Fraction of negative per-mode Lindblad torque for different disk profiles in Schwarzschild spacetime with $a_{\bullet} = 15 M$. The torques for $\pdisk = 0$ is always positive while the lower $m$-mode yield a negative torque for $\pdisk =5$. Observe that the number of modes with negative torque density increase as we lower $h_0$.}
    \label{fig:LR_torque_and_reversal}
\end{figure}

Another interesting result predicted from analytical theory was the phenomenon of torque reversal. As stated before, our calculations have not identified any torque reversals, at least for the scale heights studied in this paper. However, we do see some hints that torque reversal might still occur in ultra-thin disks, and particularly those with extreme energy density gradients. In Fig.~\ref{fig:LR_torque_and_reversal} we study the per-mode Lindblad torque in Schwarzschild spacetime with $\aEMRI = 15 M$ for $(\pdisk, \qdisk) \in \left\{ (0,0), (5,0) \right\}$. The modal Lindblad torque contributions for the constant surface density disk are always positive. However, the Lindblad torque for $\pdisk = 5$ is negative for $m \lesssim 10$ for $h_0 = 0.05$, and it increases to $m \lesssim 15$ when $h_0 = 0.03$. The total torque for $\pdisk = 5$ is ultimately positive as the positive torque contributions from the higher $m$-modes accumulate and ultimately overwhelm the negative contribution from the lower $m$-modes, leading to a net positive result.
This suggests that the number of modes with negative torque density could scale inversely with $h_0$. Hence, it might be possible to see torque reversal in very thin disks, but we leave a detailed analysis of this to the future.

\section{Conclusions}\label{sec:conclusions}
The theoretical and numerical framework presented in this works takes a significant step in advancing our understanding of how SMBH accretion disks and EMRIs affect one another. The generality and efficiency of our formulation makes it possible to calculate torques and synthesize maps of the disk response for general disk backgrounds and EMRI parameters, even in the presence of strong corotation torques. 

Our calculations identify several characteristic features of relativistic disk--EMRI interactions. Relativistic effects become significant once the EMRI reaches orbital radii of approximately $a_\bullet \lesssim 50M$, while variations in the SMBH spin can produce order-unity changes in the torque for $a_\bullet \lesssim 20M$. These effects do not lead to a universal enhancement of the torque: depending on the surface-density and sound-speed gradients, relativistic shifts of the Lindblad resonances can either increase or decrease the net angular-momentum transfer. We also find that the azimuthal mode number required to recover $95\%$ of the total torque scales approximately as $m_{95}\propto h_0^{-1}$, with $m_{95}h_0\simeq 2.0$--$3.5$ across the disk profiles considered. 

Although steep density gradients can produce negative torques in low-$m$ modes, the positive high-$m$ contribution dominates in all finite-thickness disks studied here, and we find no reversal of the total migration torque.
Our comparison with pressureless analytical models further shows that finite-pressure effects, in general, cannot be absorbed into a single adjustment of the high-$m$ torque cutoff.

One limitation of our analysis is that we retain only the leading-order contribution to the singular gravitational field of the EMRI. Incorporating higher-order terms in the local singular-field expansion is straightforward, and we expect these corrections to be subdominant, particularly for the high-$m$ modes whose resonances lie close to the perturber. Extending the calculation to include the regular, radiative part of the metric perturbation—or the full retarded metric perturbation in Lorenz gauge—is also possible in principle, but would be substantially more computationally demanding~\cite{Dyson:2026ddd}.

Since a substantial fraction of EMRIs could occur within accretion disks, the methods developed here have potential to substantially improve the analysis of gravitational wave data from future space-based detectors such as LISA. Specifically, accurate and flexible models for the influence of EMRI environments on their waveforms will make it possible to disentangle the influence of environments from potential deviations from general relativity, and to infer the physical conditions of EMRI environments. The ability of our formalism to generate surface maps of EMRI-induced disk perturbations may also assist in modeling electromagnetic counterparts to disk-embedded EMRIs.

Although in this work we have specialized to circular, non-spinning, EMRIs, the mathematical framework is general and can be used to analyze EMRIs on more general orbits. 
Some formation scenarios predict highly eccentric EMRIs and it would be interesting to extend the numerical results to analyze disk-induced torques on eccentric EMRIs, and those with spinning secondaries~\cite{Lui:2026uai}.
Finally, it would be interesting to extend our analysis beyond simple power law disk parameterization modeling the impact of realistic AGN disks on the EMRI evolution.

\section*{Acknowledgments}
We thank Adam Pound for illuminating discussions on the gravitational potential of the small object.
The authors are pleased to acknowledge that the work reported on in this paper was substantially performed using Princeton University’s Research Computing resources.
Support for this work was provided by NASA through the NASA Hubble Fellowship grant No. HST-HF2-51553.001, awarded by the Space Telescope Science Institute, which is operated by the Association of Universities for Research in Astronomy, Inc., for NASA, under contract NAS5-26555. 

\bibliography{references}

\input{appendix}
\end{document}

%% file: appendix.tex
\appendix
\section{Kerr metric}\label{appendix:Kerr-In-Plane-Coeffs}
The expressions for the metric variables in Kerr spacetime are
\begin{subequations}
\begin{align}
    \Tilde{\omega} &= 
    \frac{2 M^2 \chi }{2 M^3 \chi ^2+2 M r^2+r \Delta (r)}
    \,,\\
    \exp\left(2\nu\right) &= 
    1 - \frac{2M}{r} + \frac{2 \chi M^2}{r} \Tilde{\omega}
    \,,\\    
    \exp(2\Psi) &= 
    \frac{2 M^3 \chi ^2}{r}+2 M r+\Delta (r)
    \,,\\
    \exp(2\mu) &= \frac{r^2}{\Delta(r)}\,,\\
    \Delta &= r^2 - 2 M r +\chi^2 M^2 \,.
\end{align}
\end{subequations}

\section{Expression for $\textsf{s}_0^2$}\label{appendix:ssq-expr}
Here, we list the expression for $\textsf{s}_0^2$ in the equatorial plane for circular geodesics
\begin{widetext}
\begin{align}
    &\textsf{s}_0^2
    =
    \epssoft^2
    +
    \frac{a_{\star}^4 M^2}{(a_{\star}-2) a_{\star}+\chi^2}-\frac{2 a_{\star}^3 M r \cos (\Psi )}{(a_{\star}-2) a_{\star}+\chi^2}
    +
    \left(\frac{dt}{d\tau}\right)^2
    \frac{ r^2 \left(M \OmegaEMRI \left(a_{\star}^3+(a_{\star}+2) \chi^2\right)-2 \chi\right)^2}{2 a_{\star}^4}
    \nonumber\\
    &+\frac{1}{2} r^2 \left(\frac{(a_{\star}+2) \chi^2}{a_{\star}^3}+\frac{a_{\star}^2}{(a_{\star}-2) a_{\star}+\chi^2}+1\right)+\cos (2 \Psi ) \bigg[\frac{1}{2} r^2 \left(-\frac{(a_{\star}+2) \chi^2}{a_{\star}^3}+\frac{a_{\star}^2}{(a_{\star}-2) a_{\star}+\chi^2}-1\right)
    \nonumber\\
    &-
    \left(\frac{dt}{d\tau}\right)^2
    \frac{r^2 \left(M \OmegaEMRI \left(a_{\star}^3+(a_{\star}+2) \chi^2\right)-2 \chi\right)^2}{2 a_{\star}^4}\bigg]
    \,,
\end{align}
\end{widetext}
where
\begin{align}
    \Psi \equiv \phi - \OmegaEMRI t \,.
\end{align}

\section{Numerical Grid and Convergence}\label{appendix:grid}
In this appendix, we describe how our adaptive grid is generated, present convergence studies and discuss how our results depend on the pole displacement and softening parameters.
\subsection{Grid generation}
Our adaptive grid contains three different pieces: the left-WKB grid $r_{L,\mathrm{grid}}$, the central corotation grid $r_{C,\mathrm{grid}}$, and the right-WKB grid $r_{R,\mathrm{grid}}$. The central grid is tuned to resolve the strong corotation effects, and the left (right)-WKB grids adapt to the WKB-like waves present in regions away from the Lindblad resonances.

To generate the grid, we first need to decide the inner and outer extent of the domain for each mode. A custom grid is required for the calculation of most of the modal solutions to be efficient: lower-azimuthal-wavenumber modes typically have fairly long wavelengths and propagate throughout the entire disk; on the other hand, high-azimuthal-wavenumber modes typically have very short wavelengths, requiring a finer mesh spacing, but are localized to a narrow region around the perturber. 

Each calculation begins with the selection of broad outer and inner boundaries for the grid $r_{\mathrm{out/in}}$.
Typically, we set
\begin{align}
    r_{\mathrm{out}} = 5 \aEMRI \,,
    r_{\mathrm{in}} = r_{\mathrm{ISCO}} \,.
\end{align}
\begin{figure}[h!]
    \centering
    \includegraphics[width=\linewidth]{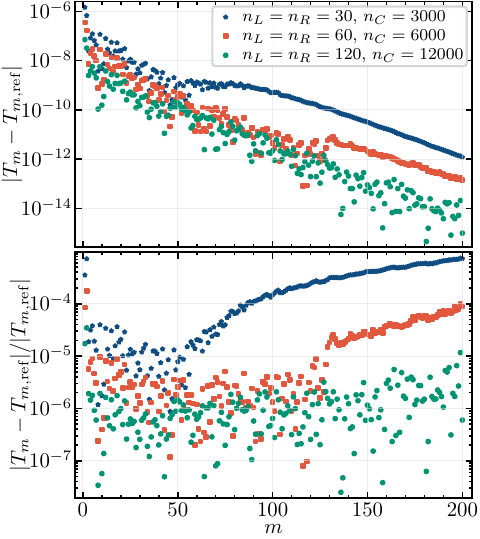}
    \caption{Convergence of the per-mode torque as functions of resolution. The top (bottom) panel shows the absolute (relative) errors from the finest resolution value. The parameters for the convergence study are: $\aEMRI = 25 M,\chi=0,\pdisk=0,\qdisk=0$ and $h_0=0.05$.}
    \label{fig:convergence}
\end{figure}
For some applications, like generating images of the wake morphology as in Figures \ref{fig:spirals} and \ref{fig:spirals_conv}, we use these as the inner and outer boundaries for the grids of each modal calculation. However, this is unnecessary for calculating torques, so we typically further refine the inner and outer boundaries for each calculation.
The physical motivation for our grid generation procedure is that once a mode has propagated sufficiently far from its inner and outer Lindblad resonances in the disk, it becomes well modeled by a WKB ansatz. Since the torque density is localized near the EMRI, it is unnecessary to solve for the modal structure very far from the resonances, since the modes and their associated torque densities die out rather quickly. 

To estimate the necessary grid extent for a given mode, we begin by generating an ancillary grid, geometrically spaced with $\sim10^4$ points between $r_{\rm in}$ and $r_{\rm out}$. On this grid, we evaluate the local WKB wavenumber $k(r)$ on this grid, and determine the points at which the WKB approximation is trustworthy. 
We define this condition heuristically by
\begin{align}\label{eq:heuristic-WKB}
    k(r)\, r > \frac{10}{h_0}.
\end{align}
Below we take the region interior to corotation as an example, but the same procedure applies to the region exterior to corotation.
We denote the closest point to corotation satisfying the above inequality as $r_{\rm WKB}$.
Starting from this point, we extend the domain away from corotation until the accumulated WKB phase satisfies
\begin{align}\label{eq:wavelength-condition}
   \int_{r'_{\rm in}}^{r_{\rm WKB}}k(r)\,dr \geq 30.
\end{align}
i.e., $15/\pi$ local wavelengths have been included in the grid. If Eq.~\eqref{eq:wavelength-condition} is reached before the fiducial boundary, the corresponding inner boundary is moved to that location. Otherwise, the original fiducial boundary is retained. We use $r_{\mathrm{in}}$ and $r_{\mathrm{out}}$ to refer to the inner and outer boundaries obtained by this procedure. We have tested larger numerical constants in Equations \ref{eq:heuristic-WKB} and \ref{eq:wavelength-condition}, and found the values above to yield converged results. 

With the inner and outer boundaries of the domain for a given mode are determined, we construct a grid between those limits upon which to solve the linear equations determining the structure of that mode. 
The central corotation grid is constructed according to 
\begin{align}
    r_{C,\mathrm{grid}} = r_c + \sigma_{w,c} \frac{\sinh(s_c x)}{\sinh(s_c)}
\end{align}
where $x$ is uniformly spaced in $[-1,1]$ with $n_C$ points.
The parameters $\sigma_{w,c}$ and $s_c$ are set to
\begin{align}
    \sigma_{w,c} = 10^{-3} r_c \,, s_c = 4 \,.
\end{align}
To facilitate smoothly matching the central region to the two outer regions, we calculate the corresponding grid spacing $\Delta r_C$ and its derivative $d(\Delta r_C)/dr$, which we use to extrapolate $\Delta r_C$ as a global \textit{trial} grid spacing $\Delta r_{\rm in}$ for the left- and right-WKB grids. The actual spacing in the left- and right-WKB grids are determined by an optimization procedure.

Let us denote the left and the right edge of the central corotation grid by $r_{\mathrm{C,in} }$ and $r_{\mathrm{C,out} }$. 
In the left-WKB (right-WKB) regions we first evaluate the local WKB wavenumber $k(r)$ over an ancillary grid spanning $[r_{\mathrm{in}}, r_{\mathrm{C,in} }]$ ($[r_{\mathrm{C, out}}, r_{\mathrm{out}}]$). 
Using the same heuristic for the trustworthiness of the WKB approximation as before, we
define [Eq.~\eqref{eq:heuristic-WKB}]
\begin{align}
    \overline{k} = \mathrm{max}[k, \, k(r = 10h_0^{-1}k^{-1})] \,.
\end{align}
We then, in each region, define a trial WKB step size 
\begin{align}
    \Delta r_{\rm WKB} = \frac{1}{\overline{k}n_W} \,,
\end{align}
where $n_W$ will be determined by an optimization procedure.

Taking the left region as an example, $n_W$ is chosen to guarantee that the final grid begins at $r_{\rm in}$ and terminates at $r_{\mathrm{C,in}}$, over the range $n_W\in[n_L,n_L+1]$ where $n_L$ is the target number of cells per unit WKB phase in the left region. The target step size on the ancillary grid is then given by
\begin{equation}
\Delta r_{L,\rm{trial}}=\mathrm{smoothMin}(\Delta r_{\mathrm{WKB,L}},\, \Delta r_{\mathrm{in}}),
\end{equation}
where we define a $\rm{smoothMin}$ function as 
\begin{align}
\rm{smoothMin}(a,b) &= \dfrac{a\exp(a/f) + b\exp(b/f)}{\exp(a/f)+\exp(b/f)}\,,\\
f&=-\max(a,b)/8\,.
\end{align}
Once this trial step size $\Delta r_{L,\rm{trial}}$ has been evaluated over the ancillary grid, we construct a cubic spline interpolant which we use to construct the left segment of the radial grid for the actual calculation, $r_{L,\mathrm{grid}}$.

The analogous procedure is applied in the right region to derive $\Delta r_R$. These grid spacing are then used to determine the cell locations for the left and right grids, $r_{L,\mathrm{grid}}$ and $r_{R,\mathrm{grid}}$, which are joined with $r_{C,\rm{grid}}$ to form the final grid. 

\begin{figure}[h!]
    \centering
    \includegraphics[width=\linewidth]{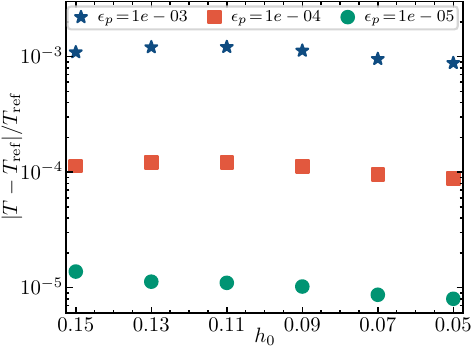}
    \caption{Convergence of the total torque as a function of the pole displacement parameter for different values of disk thickness. Our reference solution is computed at $\epsilon_p=10^{-6}$. The disk and EMRI parameters for this study are: $\pdisk =\qdisk = \chi =0$ and $a_{\bullet} = 100 M$. Observe that the relative errors scale as $\mathcal{O}(\epsilon_p)$ for all values of $h_0$ studied here. }
    \label{fig:pole_disp_errors}
\end{figure}
\subsection{Convergence}\label{appendix:convergence}
The top panel of Fig.~\ref{fig:convergence} shows the convergence rate for the per-mode torque as a function of the numerical resolution. The reference torque is calculated at $(n_L, n_R, n_C) = (240, 240, 24000)$. Observe that our results converge over the entire $m$ range. In the bottom panel, we show the relative errors. The apparent increase in the errors we see for $m>50$ in the curves is because the reference value of the torque for these high-m modes is nearly zero.

\begin{figure}[h!]
    \centering
    \includegraphics[width=\linewidth]{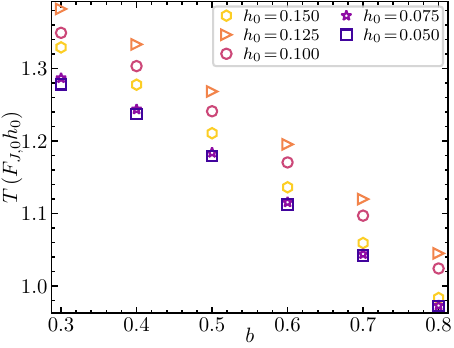}
    \caption{The dependence of the torque on the point-mass softening prescription for  $\pdisk=\qdisk=\chi=0,$ and $a_{\bullet}=100 M$. The total torque decreases as we increase the softening parameter $\bsoft$.
    }
    \label{fig:variation_with_respect_to_b}
\end{figure}

\begin{figure}[h!]
    \centering
    \includegraphics[width=\linewidth]{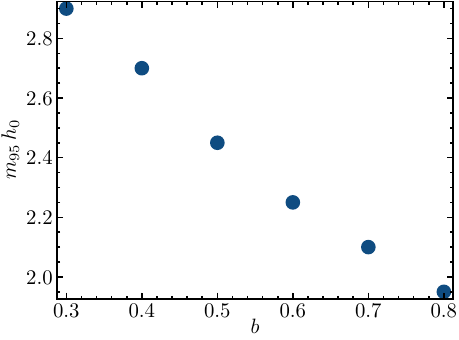}
    \caption{The dependence of $m_{95}$ on the point-mass softening parameter $\bsoft$ for $\pdisk=\qdisk=\chi=0,a_{\bullet}=30 M$ for $h_0 = 0.05$. 
    Decreasing $\bsoft$ increases the strength of the point-mass potentially, leading an increase in $m_{95}$.
    }
    \label{fig:m95_vs_b}
\end{figure}
We explore the impact of the pole displacement on the torque in Fig.~\ref{fig:pole_disp_errors}. We compute the torque for $\epsilon_{p}~\in~\left\{10^{-3}, 10^{-4}, 10^{-5}, 10^{-6} \right\}$ with $n_L=n_R=240, n_C=2400$, and compare the relative errors for each $\epsilon_p$ with respect to the smallest displacement parameter. Observe that the relative errors scale roughly as $\mathcal{O}(\epsilon_p)$ across the entire range of $h_0$ explored in this study. This shows that our corotation treatment is robust for the disk thickness considered here.

In Fig.~\ref{fig:variation_with_respect_to_b}, we analyze the impact of the point-mass softening parameter $\bsoft$ on the torque for a few different values of disk thickness. Physically, decreasing $\bsoft$ has two effects: increasing the maximum magnitude of the gravitational potential of the secondary, and adding structure to the forcing on smaller scales. Increasing $\bsoft$ then naturally decreases the torque of the EMRI on the disk. Finally, in Fig.~\ref{fig:m95_vs_b}, we study the dependence of $m_{95}$ on $\bsoft$. The decreased magnitude of the point mass potential at larger $\bsoft$ affects high$-m$ modes with resonances near the perturber but has negligible effect on low-$m$ modes; this, and the smaller scales introduced by shrinking $\bsoft$, mean that as the $\bsoft$ is decreased, more modes contribute significantly to the torque. 